\documentclass{aa}
\usepackage{graphicx}
\usepackage{txfonts}

\def\gtrsim{\mathrel{\hbox{\rlap{\hbox{\lower4pt\hbox{$\sim$}}}\hbox{$>$}}}}
\newcommand{\mincir}{\raise -2.truept\hbox{\rlap{\hbox{$\sim$}}\raise5.truept
\hbox{$<$}\ }}
\newcommand{\magcir}{\raise -2.truept\hbox{\rlap{\hbox{$\sim$}}\raise5.truept
\hbox{$>$}\ }}
\newcommand{\siml}{\raise -2.truept\hbox{\rlap{\hbox{$\sim$}}\raise5.truept
\hbox{$<$}\ }}
\newcommand{\simg}{\raise -2.truept\hbox{\rlap{\hbox{$\sim$}}\raise5.truept
\hbox{$>$}\ }}
\newcommand{\be}{\begin{equation}}
\newcommand{\ee}{\end{equation}}
\newcommand{\ba}{\begin{eqnarray}}
\newcommand{\ea}{\end{eqnarray}}
\newcommand {\h} {$h^{-1}$ Mpc $ \;$}

\newcommand {\ks} {km~s$^{-1}$ $ \;$}
\newcommand {\kss} {km~s$^{-1}$}
\newcommand {\msun} {$h^{-1} \  {\cal M}_{\odot} \;$}
\begin{document}
   \title{Morphology and luminosity segregation
of galaxies in nearby loose groups}
   \author{M. Girardi, E. Rigoni, 
F. Mardirossian and M. Mezzetti
          }

   \offprints{M. Girardi}

   \institute{
Dipartimento di Astronomia, Universit\`{a} 
degli Studi di Trieste, Via Tiepolo 11, I-34131 Trieste, Italy\\
              \email{girardi,rigoni,mardiros,mezzetti@ts.astro.it}
                          }

   \date{}

\abstract{ We study morphology and luminosity segregation of galaxies
in loose groups. We analyze the two catalogs of groups which have been
identified in the Nearby Optical Galaxy (NOG) sample, by means of
hierarchical and percolation ``friends-of-friends'' methods  (HG
and PG catalogs, respectively).  In the first part of our analysis we
consider  387 and 436 groups of HG and PG, respectively, and
compare morphology-- (luminosity--) weighted to unweighted group
properties: velocity dispersion, mean pairwise distance, and mean
groupcentric distance of member galaxies.  The second part of our
analysis is based on two ensemble systems, one for each catalog, built
by suitably combining together galaxies of all groups (1584 and
1882 galaxies for HG and PG groups, respectively).  We find that
earlier--type (brighter) galaxies are more clustered and lie closer to
the group centers, both in position and in velocity, than later--type
(fainter) galaxies.  Spatial segregations are stronger than
kinematical segregations.  These effects are generally detected at the
$\gtrsim$ 3--sigma level.  Luminosity segregation is shown to be
independent of morphology segregation.  Our main conclusions are
strengthened by the detection of segregation in both hierarchical and
percolation catalogs.  Our results agree with a continuum of
segregation properties of galaxies in systems, from low--mass groups
to massive clusters.

}
\authorrunning{Girardi et al.}
\titlerunning{Galaxy segregation in groups}
   \maketitle
%

\section{Introduction}

Groups and clusters of galaxies are complex systems involving a
variety of interacting components (galaxies, X--ray emitting gas, dark
matter).  Their investigation offers a rare opportunity to link many
aspects of astrophysics and cosmology and, in particular, to clarify
the interplay between dark and baryonic matter. 

As for galaxies, different populations, i.e. families of galaxies with
different morphology -- color -- spectral type -- luminosity, show
different distributions in projected position and LOS velocity. These
phenomena, known as segregation effects, provide a way for exploring
the connection between the distributions of dark matter and galaxies.

Segregation phenomena are well studied in galaxy clusters.  Since the
first studies (e.g., Oemler \cite{oem74}; Moss \& Dickens
\cite{mos77}; Dressler \cite{dre80}, Dressler et al.
\cite{dre97}), a long sequence of analyses has
shown that galaxies of early morphological type (red color -- low star
formation rate) are more concentrated in regions of higher projected
density and lie closer to the cluster center both in position and in
velocity than galaxies of late morphological type (blue color -- high
star formation rate), cf. Biviano et al.  (\cite{biv02}) and references
therein. Evidence for luminosity segregation is also found, although
only very luminous, possibly early--type galaxies, seem really to be
segregated from the rest of the population (e.g., Biviano et
al. \cite{biv92}; Stein \cite{ste97}; Biviano et al.  \cite{biv02}).

Observational evidence that in galaxy surveys the clustering strength
depends on morphology, luminosity, and colors (e.g., Benoist et
al. \cite{ben96}; Hermit et al. \cite{her96}, Guzzo et
al. \cite{guz97}; Norberg et al. \cite{nor01}) suggests that
segregation phenomena in galaxy systems might be connected with the
large--scale--structure formation, perhaps in the context of biased
galaxy formation or in the hierarchical growth of structure via
gravitational instability.

Alternatively and/or additionally, some environmental effects could
play an important r\^{o}le in segregation phenomena (e.g., White
\cite{whi83}; Richstone \cite{ric90}; Moss \cite{mos01}). In fact,
after a fast stage of violent relaxation, when the dynamics is
controlled by a collective potential (Lynden-Bell \cite{lyn67}),
galaxy systems should undergo a secondary relaxation phase,
characterized by a longer time scale. In this second phase, several
physical effects could modify member galaxies as regards their
internal properties, as well as their distribution in space and in
velocity.  Some of these environmental effects, such as ram pressure
stripping (Gunn \& Gott \cite{gun72}) and galaxy harassment (Moore et
al. \cite{moo96}), are less effective in group environments, where the
X--ray temperature and global potential are smaller than in
clusters. On the contrary, galaxy--galaxy interactions, such as close
tidal encounters or mergers, and dynamical friction should be
particularly significant in poor groups, where the velocity dispersion
is lower than in clusters (e.g., Sarazin \cite{sar86}; Richstone
\cite{ric90}; Makino \& Hut \cite{mak97}).

In this context, differences in the segregation effects between groups
and clusters would suggest that the system environment is fundamental
in transforming galaxies at the present epoch. On the contrary, no
difference might rather suggest that galaxy properties are influenced
by initial conditions at the time of galaxy formation, maybe through
the hierarchical growth of structures.

Unfortunately, observational difficulties prevented the researchers
from giving a precise description of segregation effects in poor
galaxy systems.  

The only strongly supported effect in groups is the spatial
segregation of galaxies of different morphological type (color --
spectral type), which is the best known effect in clusters.
Pioneering investigations were performed by Ozernoy \& Reinhard
(\cite{oze76}), Bhavsar (\cite{bha81}), and de Souza et
al. (\cite{des82}). The first systematic results using groups with
redshift information come from the study of Postman \& Geller
(\cite{pos84}). Through the analysis of groups in the CfA Redshift
Survey and in the Catalog of Nearby Galaxies they showed that the
relationship between galaxy morphology and local density, as found for
rich clusters, extends down to groups. The presence of morphological
segregation in groups of the CfA survey was also confirmed by Mezzetti
et al. (\cite{mez85}) and Giuricin et al. (\cite{giu88}), who
computed the mean groupcentric distance and the mean pairwise
distance of the members in each individual group for different galaxy
populations.  Further segregation evidence comes from the analysis of
the groups identified in the Southern Sky Redshift Survey (Maia \& da
Costa \cite{mai90}).

Recent works support the existence of morphology segregation
phenomena, too.  Mahdavi et al. (\cite{mah99}) analyzed 20 well
sampled groups finding that galaxies of different spectral types are
segregated in space.  Tran et al. (\cite{tra01}) made a deep analysis
of six X--ray detected groups finding that bulge--dominated galaxies
decreases with increasing radius, similar to the morphology--radius
relation observed in clusters.  Carlberg et al. (\cite{car01b}) analyzed
$\sim 200$ groups at intermediate redshift ($z=0.1-0.55$) identified
within the CNOC2 (Canadian Network for Observational Cosmology's field
galaxy redshift survey) and found the presence of color gradients,
i.e. the galaxies are redder toward the group center, in the most
massive groups.  Dom\'{\i}nguez et al. (\cite{dom02}) analyzed the 2dF
Group Catalog and found that the fraction related to low
star--formation galaxies depends on the local density and the
group--centric radius in the most massive ($\sim$ 40) groups.

Kinematical segregation of galaxies of different morphological  types  is
found in galaxy clusters (e.g., Stein \cite{ste97}; Biviano et
al. \cite{biv02}), but it is a smaller effect with respect to spatial
segregation.  In fact, the most different population seems to be the
10\% of galaxies with strong emission lines -- ELGs -- which have
considerably larger global velocity dispersion than other galaxy
populations (de Theije \& Katgert \cite{det99}; Biviano et
al. \cite{biv97}). The question is still open for groups: galaxies of
different spectral types seem to differ in the  value of of velocity
anisotropy, but not in the value of the global velocity dispersion
(Mahdavi et al. \cite{mah99}).

As for luminosity segregation in galaxy clusters, low levels of
significance are found both in position and in velocity. In fact, this
kind of segregation concerns only very few luminous galaxies or only
early--type galaxies (e.g., Biviano et al.  \cite{biv92}; Stein
\cite{ste97}; Adami et al. \cite{ada98}) Biviano et al. \cite{biv02}).
As for groups, the results are generally obtained by comparing
luminosity--weighted and unweighted properties of galaxy
groups. Ozernoy \& Reinhard (\cite{oze76}) found that the effect of
luminosity weighting is to increase the harmonic radius and to lower
the velocity dispersion in comparison with the unweighted
values. Further analyses of more recent group catalogs do not come on
a common conclusion.  Giuricin et al. (\cite{giu82}) claimed that the
virial parameters are largely insensitive to weighting procedures,
suggesting that group galaxies are in a status of velocity
equipartition, while Mezzetti et al. (\cite{mez85}) found that virial
radii are affected by weighting. Recent evidence for luminosity
segregation in space and in velocity comes from the study of Magtesyan
\& Movsesyan (\cite{mag95}, 303 groups), but with small significance
($\lesssim 95\%$).

In this framework, we analyze the presence of segregation effects in
galaxy groups identified in the Nearby Optical Galaxy Catalog (NOG,
Giuricin et al. \cite{giu00}, hereafter G00), which samples the local
Universe. The advantage of using NOG groups is threefold: 1) the
amount of morphological information available for nearby galaxies; 2)
the presence of many low--mass local groups in this all--sky catalog;
3) the availability of two alternative catalogs of groups identified
in the same galaxy catalog using two different algorithms. On the
other side, these groups suffer from the problem plaguing traditional
catalogs, i.e.  the poor statistics available for each group. In view
of this problem, we have devoted some effort to combining the data of
many groups.

The paper is organized as follows.  We describe the data sample and
compute the main physical group quantities in Sect.~2.  We devote
Sect.~3 and Sect.~4 to the detection and analysis of the segregation
effects.  We discuss our results and draw our conclusions in Sect.~5.

Unless otherwise stated, we give errors at the 68\% confidence level
(hereafter c.l.).

A Hubble constant of 100 $h$ \ks Mpc$^{-1}$ is used throughout.

\section{Data sample and group properties}
We analyze the galaxy loose--groups identified in the NOG catalog by G00.
NOG is a complete apparent--magnitude catalog
(corrected total blue magnitude $\rm{B}\le 14$), with an upper distance
limit (cz$<$6000 \kss), and collects $\sim 7000$ optical galaxies,
basically extracted from the Lyon--Meudon Extragalactic Database
(LEDA; c.f. Paturel et al. \cite{pat97}).  NOG covers about $2/3$ of
the sky $(|b| > 20^{\circ})$, and is quasi-complete in redshift
($97\%$).  Almost all NOG galaxies ($98.7\%$) have a morphological
classification as taken from LEDA, and parameterized by $T$ (the
morphological--type code system of RC3 catalog -- de Vaucouleurs et
al.  \cite{dev91}) with one decimal figure.

G00 identified NOG groups by means of both the hierarchical and the
percolation ``friends--of--friends'' methods. In particular, they
employed two variants of the percolation method, which gave very
similar catalogs of groups. Here we use the P2 catalog obtained with
the variant where both the distance link parameter and the velocity
link parameter scale with distance (Huchra \& Geller \cite{huc82}),
which is the most frequently used method of group identification in
the literature.  As for the hierarchical method, G00 basically
followed the procedure adopted by Gourgoulhon et al. (\cite{gou92}).

The final hierarchical and percolation catalogs contain 475 and 513
groups with at least three members, respectively (hereafter HG
and PG). The HG and PG catalogs turn out to be substantially
consistent as far as the distribution of members in groups is
concerned (see G00 for more details).

\begin{figure}
   \centering
   \includegraphics[width=8cm]{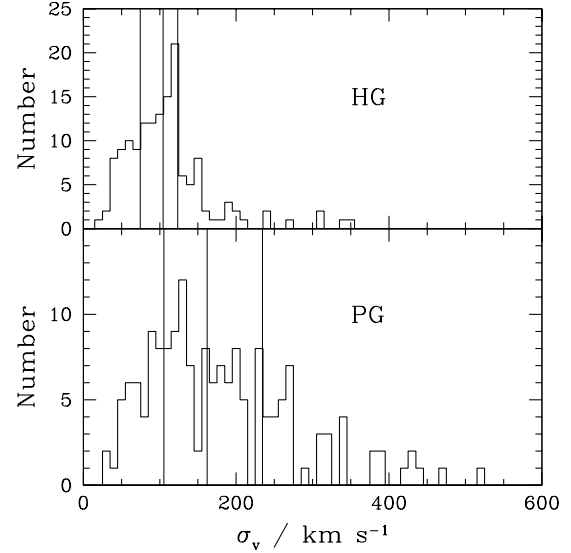}
\caption{Histograms of $\sigma_{\mathrm{v}}$--values for rich ($n\ge 5$) groups
in the HG and PG catalogs (top and bottom panels, respectively). The 
faint vertical lines indicate the separation between the quartiles of
the distribution.
\label{fig1}}
\end{figure}

The availability of these two alternative samples of groups identified
in the same galaxy catalog is of great advantage in our study.  In
fact, owing to the small number of galaxies for each group, one must
rely on members as assigned by the group-selection algorithm, rather
than using refined methods to reject interlopers as performed in
clusters or very well sampled groups (e.g., 12 groups of Zabludoff \&
Mulchaey \cite{zab98a}; 20 groups of Mahdavi et al. \cite{mah99}).  In
particular, the resulting group properties may depend on the choice of
the group-selection algorithm and its free parameters (e.g., Pisani et
al. \cite{pis92}; Frederic \cite{fre95}; Ramella et al. \cite{ram97}).
In fact, in the case of the NOG catalog, although the HG and PG groups
have a large overlap in their members, the surviving difference in
membership leads to differences in their main properties
(cf. Table~\ref{tab1} and Fig.~\ref{fig1}).  In this context, our
conclusions regarding segregation effects, where the main properties are
used in an explicit way or in rescaling parameters (cf. Sects.~3 and
4, respectively), will be strengthened by their detection in both
hierarchical and percolation catalogs.
 \begin{table*}
      \caption[]{Group properties}
         \label{tab1}

%
%

\begin{tabular}{lcccrrrrc} 
\hline \hline
\multicolumn{1}{c}{Sample}
&\multicolumn{1}{c}{$N_{\rm{GROUPs}}$} 
&\multicolumn{1}{c}{$N_{\rm{GALs}}$} 
&\multicolumn{1}{c}{$n$}
&\multicolumn{1}{c}{$z$}
&\multicolumn{1}{c}{$\sigma_{\rm{v}}$}
&\multicolumn{1}{c}{$R_{\rm{V}}$}
&\multicolumn{1}{c}{$\cal M$}
&\multicolumn{1}{c}{$R_{\rm{max}}/R_{\rm{vir}}$}
\\
\multicolumn{1}{c}{}
&\multicolumn{1}{c}{} 
&\multicolumn{1}{c}{} 
&\multicolumn{1}{c}{} 
&\multicolumn{1}{c}{}
&\multicolumn{1}{c}{km s$^-1$}
&\multicolumn{1}{c}{\h}
&\multicolumn{1}{c}{$h^{-1}\,10^{13}\,\cal M_{\sun}$}
&\multicolumn{1}{c}{}
\\
\hline
HG&387&2017&4&0.012&$89^{ +5}_{-6}$&$0.61^{ +0.04}_{ -0.05}$&$0.5^{+0.1}_{-0.1}$&3.48\\ 
HG5&148&1216&6&0.010&$104^{+8}_{-8}$&$0.65^{+0.05}_{-0.06}$&$0.7^{+0.2}_{-0.1}$&3.98\\ 
PG&436&2262&4&0.012&$138^{ +12}_{ -\phantom{0}9}$&$0.47^{+0.05}_{ -0.03}$&$0.9^{+0.2}_{-0.1}$&1.61\\
PG5&168&1375&6&0.010&$162^{+17}_{ -20}$&$0.53^{+0.04}_{ -0.04}$&$1.5^{+0.3}_{-0.4}$&1.81\\
\hline
\end{tabular}

   \end{table*}
We remove from our analysis groups with $cz\le 1000$ \ks, because when
the velocity becomes low its random component dominates and the
velocity is no longer a reliable indication of the distance.  In this
way we reject most galaxies associated with the Virgo cluster too,
i.e.  the clumps A and B, the cloud W' and the Southern Extension SE
as identified by G00 in accordance with Binggeli et al.
(\cite{bing87}) and Binggeli et al.  (\cite{bin93}). Moreover, we
remove the more distant parts of Virgo (M and W clouds), as well as
all groups identified by G00 as known clusters (cf. their Tables~5 and
7).  Obviously, we do not consider groups where the correction for
observational velocity errors (cf. below) leads to a negative value of
the velocity dispersion of member galaxies. Our final samples, HG and
PG, contain 387 and 436 groups with at least three members for
a total of 2017 and 2262 galaxies, respectively.  Of the HG
groups, 350 comprise $n<10$ members (in particular 155 have $n=3$
members), 31 groups comprise $10\le n<20$ members, and 6 groups have
$n\ge 20$ members.  Of the PG groups, 394 comprise $n<10$ members
(in particular 185 have $n=3$ members), 36 groups comprise $10\le
n<20$ members, and 6 groups have $n\ge 20$ members.

Out of the above groups, 148 and 168 groups have $n\ge5$ members for a
total of 1216 and 1375 galaxies (HG5 and PG5, respectively). The
physical reality of very poor groups has often been discussed in the
literature.  In particular, the efficiency of the percolation
algorithm has been repeatedly checked, showing that an appreciable
fraction of the poorer groups, those with $n<5$ members, might be
false (i.e. represent unbound density fluctuations), whereas the
richer groups almost always correspond to real systems (e.g., Ramella
et al. \cite{ram89}; Ramella et al. \cite{ram95}; Mahdavi et
al. \cite{mah97}; Nolthenius et al. \cite{nol97}; Diaferio et
al. \cite{dia99}). Throughout the paper we apply our analysis
to these statistically more reliable groups, too.

We take from G00 the data available for galaxy positions, redshifts
(in the Local Group rest frame), corrected total blue magnitudes, and
morphologies.  For each group, we compute the main physical
quantities.

We calculate the mean group velocity $\bar{\rm{v}}$ by using the
biweight estimator (Beers et al. \cite{bee90}).  We also compute the
biweight group center by the mean of member positions, , i.e. we
perform the biweight mean-estimator of member right-ascensions and
declinations, separately.  The group size is then given by
$R_{\rm{max}}$, which is the (projected) distance of the most distant
galaxy from the group center.

The main property of a galaxy system is the LOS velocity dispersion,
$\sigma_{\mathrm{v}}$.  In fact, the virial radius
$R_{\mathrm{vir}}$, which defines the region where the matter
overdensity is $\sim 178$ (for a CDM $\Omega_{\mathrm{m}}=1$ cosmology),
scales with $\sigma_{\mathrm{v}}$, and the mass in the
virialized region $\cal M$ ($<R_{\mathrm{vir}}$) scales with
$\sigma_{\mathrm{v}}^3$ (e.g., Carlberg et al. \cite{car97};
Girardi et al.  \cite{gir98}; Antonuccio-Delogu et al. \cite{ant02}).
We estimate $\sigma_{\mathrm{v}}$ by using the biweight
estimator for groups with $n\ge15$ and the gapper estimator for groups
with $n<15$ (Beers et al. \cite{bee90}).  We apply the
relativistic correction and the usual correction for velocity errors
(Danese et al. \cite{dan80}).  In particular, for each galaxy, we
assume a typical velocity error of $30$ \ks based on the average
errors estimated in the RC3 catalog from optical and radio
spectroscopy (de Vaucouleurs et al.  \cite{dev91}).

As for $R_{\mathrm{vir}}$, we adopt the definition by Girardi et
al. (\cite{gir98}), recovered by using the observational King-modified
galaxy distribution of nearby clusters:
\begin{equation}
R_{\mathrm{vir}}=[2\cdot\sigma_{\mathrm{v}}/(1000~\mbox{km}~\mbox{s}^{-1})]~h^{-1}~\mbox{Mpc}.
\end{equation}
\noindent 
The use of an isothermal distribution leads to a similar value
(cf. our $R_{\mathrm{vir}}$ with the $R_{200}$ definition by Carlberg et al.
\cite{car97}).

Another important group quantity is the projected radius $R_{\rm{V}}$,
which is twice the harmonic radius and is often used in the
computation of the virial mass (e.g., Binney \& Tremaine \cite{bin87};
Girardi et al. \cite{gir98}):
\begin{equation}
{\cal M}(<R_{\mathrm{max}})=\frac{3\rm{\pi}}{2} \cdot \frac{R_{\mathrm{V}} \sigma_{\mathrm{v}}^2}{G},
\end{equation}
\noindent where the factor $3\rm{\pi}/2$ is the deprojection factor
adequate for spherical systems, and the mass is that contained within
the sampled region, i.e. for $R\le R_{\mathrm{max}}$.  This
approach to mass estimate is similar to that computed in other group
catalogs (e.g., Ramella et al. \cite{ram02}, and refs. therein), see
also Girardi \& Giuricin (\cite{gir00}) for related discussions.

The radius $R_{\mathrm{V}}$ is defined as:
\noindent
\begin{equation}
R_{\mathrm{V}}=n(n-1)/\sum_{i>j} R_{ij}^{-1},
\end{equation}
\noindent where $R_{ij}$ are the projected mutual galaxy distances,
and $n$ is the number of group members within the sampled region.  The
value of $R_{\mathrm{V}}$ depends on the relative galaxy distribution, and 
the extension of the sampled region, i.e. the region occupied by the
galaxies used in the computation. 

Table~\ref{tab1} lists the median values  (and $90\%$ confidence
intervals) of the above group properties for the HG and PG catalogs.
The confidence intervals are computed following the
procedure\footnote{For the median of an ordered distribution of
$N$ values the confidence intervals $x_{(r)}$ and $x_{(N-r+1)}$,
corresponding to a probability $P(x_{(r)}\le x\le
x_{(N-r+1)})=1-\alpha$, can be obtained from $1-\alpha=2^{-N}
\sum^{N-r}_{i=r} \left({N}\atop {i}\right)$.}  described by Kendall
\& Stuart (\cite{ken79}, eq. 32.23) and first proposed by Thompson
(\cite{tho36}).  Values for groups with $n\ge5$ are shown, too.

Median values of $R_{\mathrm{V}}$ are comparable to typical values
quoted in the literature (cf. Table~4 of Tucker et al. \cite{tuc00},
-- note that their deprojected $R_{\mathrm{h}}$ corresponds to
$\mathrm{\pi}/4\times R_{\mathrm{V}}$).

As for $\sigma_{\mathrm{v}}$, the HG value is $\sim 2/3$ of the PG
value (cf. also Pisani et al. \cite{pis92} for a similar
result). Indeed, it has been suggested that the drawback of
percolation methods is the inclusion in the catalogs of possible
non--physical systems, like a long galaxy filament aligned close to
the line of sight, which give large velocity-dispersion estimates,
while the drawback of hierarchical methods is the splitting of galaxy
clusters into various subunits, which give small velocity--dispersion
estimates (e.g., Gourgoulhon et al. \cite{gou92}). The resulting
difference in the distribution of velocity--dispersion estimates is
outlined in Fig.~\ref{fig1}.  Our value of PG $\sigma_{\mathrm{v}}$
lies within the range of values reported by Tucker et
al. (\cite{tuc00}), which are computed for percolation catalogs, and
that of HG $\sigma_{\mathrm{v}}$ is comparable with results for the
hierarchical catalog of Tully (\cite{tul87}).

Note that NOG groups are sampled well outside $R_{\mathrm{vir}}$, in agreement
with the conclusion of Girardi \& Giuricin (\cite{gir00}) about groups
by Garcia (\cite{gar93}). This characteristic is common in traditional
catalogs of loose groups and in fact, Carlberg et al. (\cite{car01a})
introduced a variant of the friends--of--friends algorithm in such a
way as to consider only central, possibly virialized regions of CNOC
groups. In Sect.~4 we will consider only central group regions, too.

\section{Detection of segregation}

To detect possible segregation effects in velocity we compare several
estimates of velocity dispersions (cf. also Biviano et
al. \cite{biv92}).  The unweighted velocity dispersion is defined as:

\begin{equation}
\sigma_u = \left[\frac{\sum_{i=1}^{n} (\rm{v}_{i} - \bar{\rm{v}})^{2}}{n-1}
- \Delta \right]^{1/2},
\end{equation}

\noindent where $\rm{v}_{i}$ is the velocity of the i-th galaxy corrected for
cosmological effects, $\bar{\rm{v}}=\sum_{i=1}^{n} \rm{v}_{i}/n$, and
$\Delta$ is the correction term which accounts for the measurement
errors (see, e.g., Danese et al. \cite{dan80}).

To check for possible luminosity segregation, we use
the luminosity-weighted velocity dispersion:
\begin{equation}  
\sigma_{lw} = \left[\frac{\sum_{i=1}^{n} (\rm{v}_{i} - \bar{\rm{v}})^{2} \, l_{i}}
{\sum_{i=1}^{n} l_{i}} \, \frac{n}{n-1} - \Delta \right]^{1/2},
\end{equation}
\noindent where 
$l_{i}$ is the apparent galaxy blue luminosity;
the groupcentric-distance--weighted velocity
dispersion:
\begin{equation}
\sigma_{dw} = \left[\frac{\sum_{i=1}^{n} (\rm{v}_{i} - \bar{\rm{v}})^{2}
R_{i}^{-1}}{\sum_{i=1}^{n} R_{i}^{-1}} \, \frac{n}{n-1} - \Delta
\right]^{1/2},
\end{equation}
\noindent 
where $R_{i}$ is the projected galaxy distance from its group--center;
and the luminosity--groupcentric-distance--weighted velocity dispersion:
\begin{equation}
\sigma_{ldw} = \left[\frac{\sum_{i=1}^{n} (\rm{v}_{i} - \bar{\rm{v}})^{2} \, l_{i} R_{i}^{-1}}{\sum_{i=1}^{n} l_{i} 
\, R_{i}^{-1}} \, \frac{n}{n-1} - \Delta \right]^{1/2},
\end{equation}
\noindent where the effects of luminosity-- and 
groupcentric-distance--weighting are coupled.

\noindent Moreover, we introduce the morphology-weighted velocity
dispersion, $\sigma_{Tw}$:
\begin{equation}  
\sigma_{Tw} = \left[\frac{\sum_{i=1}^{n} (\rm{v}_{i} - \bar{\rm{v}})^{2} \, k_{i}}
{\sum_{i=1}^{n} k_{i}} \, \frac{n}{n-1} - \Delta \right]^{1/2},
\end{equation}
\noindent where $k_i$ is a weight related to the galaxy morphology:
$k=4$ for early--type galaxies and $k=1$ for late--type galaxies,
i.e. galaxies with $T<0$ and with $0\le T<10$.  The choice of a factor
of four of difference in the weights is suggested by the typical
difference in $M/L_B$ ratios for early-- and late--type galaxies
(e.g., Bahcall et al. \cite{bah95}).

To detect possible spatial segregation effects, we consider the
(projected) mean distance of members from the center of the group:
\begin{equation}
\bar{R}=\sum_{i=1}^{n} R_i/n,
\end{equation}
\noindent and the
(projected) mean pairwise separation of the members:
\begin{equation}
\bar{R}_{ij}=\frac{\sum_{i>j} R_{ij}}{n(n-1)/2};
\end{equation}
\noindent and the respective weighted quantities:
\begin{equation}
\bar{R}_{lw}=\sum_{i=1}^{n} l_i R_i/\sum_{i=1}^{n}l_i, 
\end{equation}
\begin{equation}
\bar{R}_{Tw}=\sum_{i=1}^{n} k_i R_i/\sum_{i=1}^{n}k_i,
\end{equation}
\begin{equation}
\bar{R}_{ij,lw}=\frac{\sum_{i>j} l_i l_j R_{ij}}{\sum_{i>j} l_i l_j},
\end{equation}
\begin{equation}
\bar{R}_{ij,Tw}=\frac{\sum_{i>j} k_i k_j R_{ij}}{\sum_{i>j} k_i k_j}.
\end{equation}

We use the {\em Sign} and {\em Wilcoxon Signed-ranks} tests (hereafter
referred to as {\em S}-- and {\em W}--tests, e.g. Siegel \cite{sie56}) to
compare weighed and unweighted estimates of each quantity we compute
below.  {\em S}-- and {\em W}--tests are nonparametric tests which
investigate the median difference between pairs of scores from two
matched sample of a certain size. The {\em W}--test differs from the
{\em S}--test in that the magnitude of score differences within pairs is
taken into account, rather than simply the direction of such
differences.

We find that each weighted quantity is smaller than the respective
unweighted quantity.  The {\em S}-- and {\em W}--tests recover a
strong significant difference with the exception of $\sigma_{Tw}$, for
which only a partial significance is generally found
(cf. Table~\ref{tab2}).  These results agree with the scenario where
earlier--type (brighter) galaxies are more clustered and lie closer to
the group center both in position and in velocity than later--type
(fainter) galaxies. Moreover, galaxies which are close to the group
center move slowly.

These results are confirmed when only statistically more reliable
groups are considered ($n\ge 5$ members, HG5 and PG5 samples).

 \begin{table*}
      \caption[]{Comparison of weighted and unweighted group properties}
         \label{tab2}
%
%
\begin{tabular}{lrrrrrrrr}
\hline \hline
\multicolumn{1}{c}{}
&\multicolumn{2}{c}{HG: 387 systems}
&\multicolumn{2}{c}{HG5: 148 systems}
&\multicolumn{2}{c}{PG: 436 system}
&\multicolumn{2}{c}{PG5: 168 systems}
\\
\multicolumn{1}{c}{X--Y}
&\multicolumn{1}{c}{$P_S$}
&\multicolumn{1}{c}{$P_W$}
&\multicolumn{1}{c}{$P_S$} 
&\multicolumn{1}{c}{$P_W$}
&\multicolumn{1}{c}{$P_S$} 
&\multicolumn{1}{c}{$P_W$}
&\multicolumn{1}{c}{$P_S$} 
&\multicolumn{1}{c}{$P_W$}
\\
\hline
$\sigma_{lw}$--$\sigma_u$        
&$>$99.9&$>$99.9&$>$99.9&$>$99.9&$>$99.9&$>$99.9&$>$99.9&$>$99.9\\
$\sigma_{dw}$--$\sigma_u$        &$>$99.9&$>$99.9&   98.2&   
98.8&99.9&$>$99.9&   98.2&$>$99.9\\ 
$\sigma_{ldw}$--$\sigma_u$       
&$>$99.9&$>$99.9&$>$99.9&$>$99.9&$>$99.9&$>$99.9&99.6&$>$99.9\\ 
$\sigma_{Tw}$--$\sigma_u$        &   83.8&   95.0&   97.4&   
99.0&   92.8&   96.7&   97.1&   96.0\\ 
$\bar{R}_{lw}$--$\bar{R}$          
&$>$99.9&$>$99.9&$>$99.9&$>$99.9&$>$99.9&$>$99.9&$>$99.9&$>$99.9\\ 
$\bar{R}_{Tw}$--$\bar{R}$          
&$>$99.9&$>$99,9&$>$99.9&$>$99.9&$>$99.9&$>$99.9&$>$99.9&$>$99.9\\
$\bar{R}_{ij,lw}$--$\bar{R}_{\rm{ij}}$  
&$>$99.9&$>$99.9&$>$99.9&$>$99.9&$>$99.9&$>$99.9&$>$99.9&$>$99.9\\    
$\bar{R}_{ij,Tw}$--$\bar{R}_{\rm{ij}}$  
&$>$99.9&$>$99.9&$>$99.9&$>$99.9&$>$99.9&$>$99.9&$>$99.9&$>$99.9\\  
\hline 
\end{tabular}

   \end{table*}

Table~\ref{tab2} lists each comparison between weighted and unweighted
quantities and gives the percent significance of the difference according to
the {\em S}-- and {\em W}--test ($P_S$ and $P_W$, respectively).

From a quantitative point of view: the typical amount of the
difference between weighted and unweighted quantities is small, at
most $7\%$ (in the case of $\sigma_{ldw}$ vs. $\sigma_{\rm {u}}$).
Thus, any segregation effect has a little relevance in the computation
of global group quantities.  On the contrary, we stress that, as
regards the connection between galaxy evolution and environment, the
r\^{o}le of the above segregation effects is very important and could
help to clarify many points.  This explains the attempt of the next
Section and previous analyses to stack together many groups.

\section{Analysis of segregation}

\subsection{Ensemble groups}

Here we analyze the behavior of relations between interesting galaxy
properties.  Unfortunately, owing to the small number of group
members, we cannot address this question by analyzing each individual
group. Therefore, we build two ensemble systems, one for each examined
catalog, HG and PG, by combining together galaxies of all groups.  The
procedure of stacking groups of various sizes and masses into an
ensemble system requires that individual galaxy quantities are
properly scaled.

The magnitudes are normalized to the magnitude of the third-ranked
galaxy, $m_3$ in each group by using $m-m_3$ (cf. also Biviano et al.
\cite{biv92}).  Here the median value of $m_3$ corresponds to a blue
corrected absolute magnitude of $M_B \sim -19.5+5\,$log$\,h$.  We use
this normalization to take into account possible biases introduced by
the intrinsic nature of the NOG group catalog. In fact, members of
more distant groups identified in an apparent magnitude galaxy survey
are, on average, more intrinsically luminous.

The rest frame LOS velocities are normalized to the global value
of velocity dispersion of each group. For each galaxy we consider the
absolute quantity $|\rm{v}-\bar{\rm{v}}|/\sigma_{\mathrm{v}}$. Note
that the standard estimate $\sigma_u$ strongly correlates with the
robust estimate $\sigma_{\mathrm{v}}$ (at the $>99.9\%$ c.l.); thus
the results are not affected by the particular choice of
velocity--dispersion estimate.

In studies of galaxy clusters, projected groupcentric-distances $R$
are generally rescaled with $R_{\mathrm{vir}}$, whose estimate is
proportional to $\sigma_{\mathrm{v}}$ (cf. Sect.~2).  However, the
question is less obvious for galaxy groups.

Mahdavi et al. (\cite{mah99}) pointed out that, in the case of poor
systems, the possible source of errors on the observationally
estimated $R_{\mathrm{vir}}$ could be much larger than in the case of clusters.
In fact, uncertainties typical of groups, such as their dynamical
status and galaxy distribution, add to large uncertainties in the
estimate of velocity dispersions, connected to the small number of
group members. 

Moreover, owing to their different nature, distance and
velocity variables are treated in different ways in the group
identification algorithms. Thus, if the use of $\sigma_{\mathrm{v}}$
to rescale velocities is fully self--consistent, the use of a quantity
proportional to $\sigma_{\mathrm{v}}$ might be not the best choice for
rescaling distances.

To further investigate the question, we consider three possible
alternative factors for rescaling $R$, i.e. using:
1)~$R_{\mathrm{vir}}$, 2)~$R_{\rm{V}}$, and 3)~no rescaling at all. We
start from the theoretical prejudice that halo scale--invariance
implies that the existence of very different kinds of galaxy
distributions for low-- and high--$\sigma_{\mathrm{v}}$ groups is an
indication of a non--corrected scaling factor.  Thus we consider the
ensemble systems constructed from low-- and
high--$\sigma_{\mathrm{v}}$ groups (according to the median value of
$\sigma_{\mathrm{v}}$ in Table~\ref{tab1}) and we compare the two
cumulative distributions of normalized groupcentric-distances through
the Kolmogorov--Smirnov test ({\em K--S} test, e.g. Lederman
\cite{led82}): they are always different, but the smallest amount of
difference is found when using $R_{\rm{V}}$ (according to the
statistic D of the {\em K--S} test).

Thus, we choose to rescale groupcentric-distances with
$R_{\rm{V}}$. However, we have verified that our main results are
still valid in the case of the other two choices of normalization.

   \begin{figure}
   \centering
   \includegraphics[width=8cm]{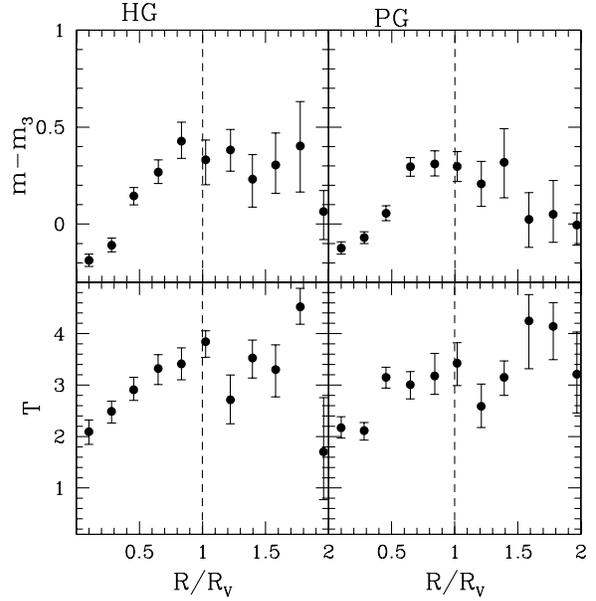}
\caption{Results for ensemble groups: (normalized) magnitude $m-m_3$,
and morphological--type $T$ vs. (normalized) projected group--centric
distance $R/R_{\rm{V}}$, respectively from top to bottom.  Points are
biweight mean values with 68\% bootstrap error--bars.  The dashed
vertical lines indicate $1\times R_{\rm{V}}$, the external limit for
following analyses, which corresponds to $\sim 2$--3$\times R_{\mathrm{vir}}$
and contains $\sim 70$--$80\%$ of all galaxies.  This plot is shown to
provide a first look at the original data: note that relations
involving magnitudes should be properly corrected
(cf. Figs.~\ref{fig3} and \ref{fig4} and Sect.~4.2).  } \label{fig2}
\end{figure}

Fig.~\ref{fig2} shows the behavior of magnitude, and
morphological--type vs.  projected groupcentric distance.  To be
conservative, hereafter we consider only the group region within one
$R_{\rm{V}}$, which contains enough galaxies to work with a large
statistical data--base: $\sim 70$--$80\%$ of the whole sample,
i.e. 1584 and 1882 galaxies for the HG and PG catalogs, respectively.
This radius corresponds to $\sim 2$--3~$\times R_{\mathrm{vir}}$ for
PG and HG respectively, and thus is not too much larger than the
virialized region, i.e. where secondary relaxation--processes of some
importance might operate.

\subsection{Segregation effects}

We consider the resulting relations for the two ensemble groups, HG
and PG: the groupcentric-distance--magnitude relation ($R$--$M$); the
groupcentric-distance--morphological type relation ($R$--$T$); the
velocity--magnitude relation ($V$--$M$); and the
velocity--morphological type relation ($V$--$T$).  We analyze
segregation effects in the framework of the Kendall rank correlation
analysis, which is completely nonparametric (Kendall \cite{ken48};
cf. also Siegel \cite{sie56}).

For each interesting relation $x$~vs.~$y$, we compute the Kendall
coefficient, $K_{xy}$, and the respective significance of the
correlation, $P_{xy}$. 

When analyzing a correlation, there is always the possibility that
this correlation is due to the association between each of the two
physical properties and a third property connected to the selection
effects of the catalog.  Statistically, this problem may be attacked
by methods of partial correlation.  In partial correlation, the
effects of variation by a third $z$ variable upon the relation between
the $x$ and $y$ variables are eliminated.  In practice, the
correlation between the two interesting properties is found with the
third variable kept constant.  Thus we also compute the Kendall
partial correlation coefficient, $K_{xy,z}$, which is a measure of the
correlation between two data sets, $x$ and $y$, independently of their
correlation with a third data set, $z$ (Kendall \cite{ken48};
cf. eq. 9.13 of Siegel \cite{sie56}):
\begin{equation}
K_{xy,z}=\frac{K_{xy}-K_{zy}K_{xz}}{\sqrt{(1-K_{zy}^2)(1-K_{xz}^2)}},
\end{equation}
\noindent 
where, as for the third variable, we consider both the group distance
($\propto \bar{\rm{v}}$), and the number of group--members (richness
$n$).  In fact, the distance is connected with the selection function
of the group catalog, which is based on an apparent magnitude complete
galaxy sample.  Moreover, in such very poor systems ($n\gtrsim 3$) the
normalized magnitude $m-m_3$ strongly correlates with $n$, which in
its turn is slightly correlated with both size and velocity dispersion
(cf. Table~\ref{tab1}).

 \begin{table*}
      \caption[]{Results for ensemble groups}
         \label{tab3}

%
%
\begin{tabular}{ccrrrr}
\hline \hline
\multicolumn{1}{c}{Sample}
&\multicolumn{1}{c}{$x$~vs.~$y$}
&\multicolumn{1}{c}{$N_{\rm{GALs}}$}
&\multicolumn{1}{c}{$K_{xy}$, $P_{xy}$} 
&\multicolumn{1}{c}{$K_{xy,\bar{\rm{v}}}$, $P_{xy,\bar{\rm{v}}}$} 
&\multicolumn{1}{c}{$K_{xy,n}$, $P_{xy,n}$}
\\
\hline
HG Groups&$R$--$M$&1584& 0.19, $>$99.9 & 0.19, $>$99.9  & 0.13, $>$99.9 \\
         &$R$--$T$&1558& 0.07, $>$99.9 & 0.07, $>$99.9  & 0.09, $>$99.9 \\
         &$V$--$M$&1584& 0.06, $>$99.9 & 0.05, $>$99.9  & 0.04, 
$\phantom{>} 99.6$\\
         &$V$--$T$&1558& 0.02, $\phantom{>} 84.7$& 0.02, 
$\phantom{>} 93.0$ & 0.02, $\ 
89.3$\\
PG Groups&$R$--$M$&1882& 0.16, $>$99.9 & 0.16, $>$99.9  & 0.09, $>$99.9 \\
         &$R$--$T$&1856& 0.06, $>$99.9 & 0.06, $>$99.9  & 0.06, $>$99.9 \\
         &$V$--$M$&1882& 0.06, $>$99.9 & 0.06, $>$99.9  & 0.03, 
$\phantom{>} 99.3$\\
         &$V$--$T$&1856&$-0.01$, $\ 67.1$&$-0.01$, $\phantom{>} 
69.0$ & 
0.00, $\phantom{>} 54.4$\\ \hline 
\end{tabular}

   \end{table*}

For both the HG and PG catalogs, Table~\ref{tab3} shows the results
for the ensemble groups constructed as described in Sect.~4.1.  For
each interesting relation $x$~vs.~$y$, we give the Kendall
coefficient, $K_{xy}$, and the Kendall partial correlation coefficient,
which takes into account the effect of the third variable,
$K_{xy,\bar{\rm{v}}}$ and $K_{xy,n}$.  We also list the respective
significance of the correlations, $P$.  The value of $P_{xy}$ is
recovered form the fact that the sampling distribution of $K_{xy}$ is
practically indistinguishable from the normal distribution (Kendall
\cite{ken48}; cf. eq. 9.11 of Siegel \cite{sie56}). Since the sampling
distribution of $K_{xy,z}$ is unknown, to compute $P_{xy,\bar{\rm{v}}}$
and $P_{xy,n}$ we adopt the bootstrap method performing 1000
bootstrap resamplings for each correlation.

The Kendall coefficients of the $R$--$T$ and $V$--$T$ correlations
show no decrease when the $\bar{\rm{v}}$ or $n$ variables are taken
into account ($K_{xy,\bar{\rm{v}}}\sim K_{xy,n}\sim K_{xy}$).
Therefore the $R$--$T$ and $V$--$T$ relations are not biased by
systematic effects.

As for the $R$--$M$ and $V$--$M$ relations, we find that
$K_{xy,\bar{\rm{v}}}\sim K_{xy}$, while $K_{xy,n}$ is systematically
smaller than $K_{xy}$ although still giving significant correlation.
Thus, we conclude that the $R$--$M$ and $V$--$M$ correlations are due
to a combination of a true physical effect and a spurious one.  The
spurious one is due to our stacking of groups with different richness.

To support the above analysis, we also make Monte Carlo simulations
performing a random shuffle of galaxy velocities, magnitudes, and
morphological types within each individual group. This procedure,
which leaves unchanged global group properties and modifies the group
internal structure, should destroy any physical segregation effects
and reveal the presence of spurious correlations.  We construct
the ``simulated'' ensemble--group by combining together the galaxies
of groups resulting from Monte Carlo simulations.  In particular, we
perform 15 Monte Carlo simulations for each group in order
to consider a large ($>20000$) number of galaxies and thus stabilize
the resulting Kendall coefficient, $K_{xy}^{\rm{sim}}$.  No
significant correlation is found for $R$--$T$ and $V$--$T$
relations. When considering $R$--$M$ and $V$--$M$, the simulated group
shows a significant correlation, characterized by a
$K_{xy}^{\rm{sim}}$ smaller than the coefficient of the original
data $K_{xy}$.  Moreover, when using only groups with a fixed number
of members $n=$3, 4, 5, and 6 (groups with $n>6$ are statically poorly
represented) the corresponding Monte Carlo simulated groups show no
correlation at all. Thus, in agreement with Kendall partial
correlation analysis, we conclude that $R$--$M$ and $V$--$M$
correlations are partially spurious due to our stacking of groups with
different richness.

Hereafter, to take into accounts the richness effect, we consider as
really physical meaningful the $K_{xy,n}$ coefficient.

Having addressed the problem of possible biases, we can give our
results about segregation effects, which are similar for both the HG
and PG catalogs.  Spatial correlations, in luminosity or morphology
($R$--$M$, $R$--$T$), are the strongest one ($K_{xy,n}=0.06$-$0.13$).
Luminosity segregation in velocity ($V$--$M$) is characterized by a
small correlation coefficient ($K_{xy,n}=0.03$-$0.04$), still
significant.  Morphological segregation in velocity ($V$--$T$) is not
significant, in agreement with the fact that this is the faintest
segregation detected in Sect.~3.

   \begin{figure}
   \centering
   \includegraphics[width=8cm]{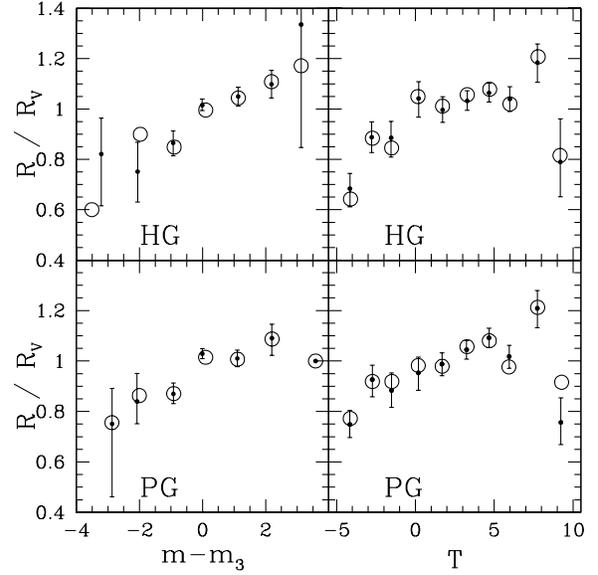}
\caption{Groupcentric-distance vs. magnitude, and morphological type:
$R$--$M$ and $R$--$T$ relations in the left and right panels,
respectively.  Points are biweight mean values for all groups (filled
circles) and rich $n\ge 5$ groups only (open circles).  Error bars are
68\% bootstrap estimates. 
For the sake of clarity, error bars are
shown for one sample only.
Observational results are normalized point
by point with results from simulated groups (cf. text).
\label{fig3}}
\end{figure}

   \begin{figure}
   \centering
   \includegraphics[width=8cm]{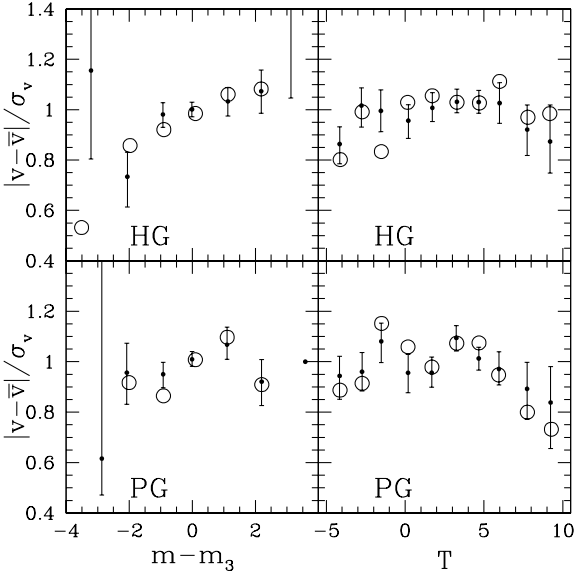}
\caption{Velocity vs. magnitude, and morphological type: $V$--$M$ and
$V$--$T$ relations in the left and right panels, respectively.  Points
are biweight mean values for all groups (filled circles) and rich
$n\ge 5$ groups only (open circles). Error bars are 68\% bootstrap
estimates.  For the sake of clarity, error bars are shown for one
sample only.  Observational results are normalized point by point with
results from simulated groups (cf. text).
\label{fig4}}
\end{figure}

The interesting relations are better visualized in Figs.~\ref{fig3}
and \ref{fig4} for both all and rich $n\ge 5$ groups. There, to
take into account the spurious component in the $R$--$M$ and $V$--$M$
relations, we show our results rescaled to those of the corresponding
simulated group, i.e.  we normalize the (y-axis) values obtained
for the real ensemble-group to the values obtained for the
corresponding simulated-group with the same binning procedure.  For
the sake of homogeneity, we apply the same procedure to $R$--$T$ and
$V$--$T$ relations.

\subsection{Segregation and group richness}

We analyze separately groups with $n<5$ and those with $n\ge5$
members. Tables~\ref{tab4} and \ref{tab5} show the results of the
correlation analysis.

 \begin{table}
      \caption[]{Poor ($n<5$) groups}
         \label{tab4}
%
%
\begin{tabular}{ccrr}
\hline \hline
\multicolumn{1}{c}{Sample}
&\multicolumn{1}{c}{$x$~vs.~$y$}
&\multicolumn{1}{c}{$N_{\rm{GALs}}$} 
&\multicolumn{1}{c}{$K_{xy,n}$, $P_{xy,n}$}
\\
\hline
HG&$R$--$M$& 577&0.10, $>$99.9\\
  &$R$--$T$& 565&0.07, $\phantom{>} 99.4$\\
  &$V$--$M$& 577&0.00, $\phantom{>} 52.7$\\
  &$V$--$T$& 565&$-0.02$,$\phantom{>} 76.7$\\
PG&$R$--$M$& 705&0.06, $\phantom{>} 99.1$\\
  &$R$--$T$& 693&0.04, $\phantom{>} 95.9$\\
  &$V$--$M$& 705&0.02, $\phantom{>} 77.3$\\
  &$V$--$T$& 693&0.01, $\phantom{>} 62.5$\\
\hline
\end{tabular}

   \end{table}
 \begin{table*}
      \caption[]{Rich ($n\ge5$) groups}
         \label{tab5}
%
%
\begin{tabular}{ccrrrr}
\hline \hline
\multicolumn{1}{c}{Sample}
&\multicolumn{1}{c}{$x$~vs.~$y$}
&\multicolumn{1}{c}{$N_{\rm{GALs}}$} 
&\multicolumn{1}{c}{$K_{xy,n}$, $P_{xy,n}$}
&\multicolumn{1}{c}{$K_{xy,n,m-m3}$, $P_{xy,n,m-m_3}$}
&\multicolumn{1}{c}{$K_{xy,n,T}$, $P_{xy,n,T}$}
\\
\hline
HG&$R$--$M$&1007&0.13, $>$99.9&           &0.12, $>$99.9\\
  &$R$--$T$& 993&0.10, $>$99.9&0.09, $>$99.9&           \\
  &$V$--$M$&1007&0.06, $\phantom{>} 99.9$&           &0.05, $\phantom{>} 99.7$\\
  &$V$--$T$& 993&0.05, $\phantom{>} 99.4$&0.05, $\phantom{>} 98.6$   &           
\\
PG&$R$--$M$&1177&0.09, $>$99.9&           &0.08, $>$99.9\\
  &$R$--$T$&1163&0.10, $>$99.9&0.09, $>$99.9&           \\
  &$V$--$M$&1177&0.03, $\phantom{>} 92.4$   &           &0.03, $\ 93.1$   
\\
  &$V$--$T$&1163&0.00, $\phantom{>} 58.5$&$-0.01$, $\phantom{>} 64.7$  &           
\\
\hline
\end{tabular}

   \end{table*}

The analysis of rich groups confirms the correlations found for the
whole sample. In particular, the $V$-$T$ correlation is now significant in
the case of the HG catalog.
 
   \begin{figure}
   \centering
   \includegraphics[width=8cm]{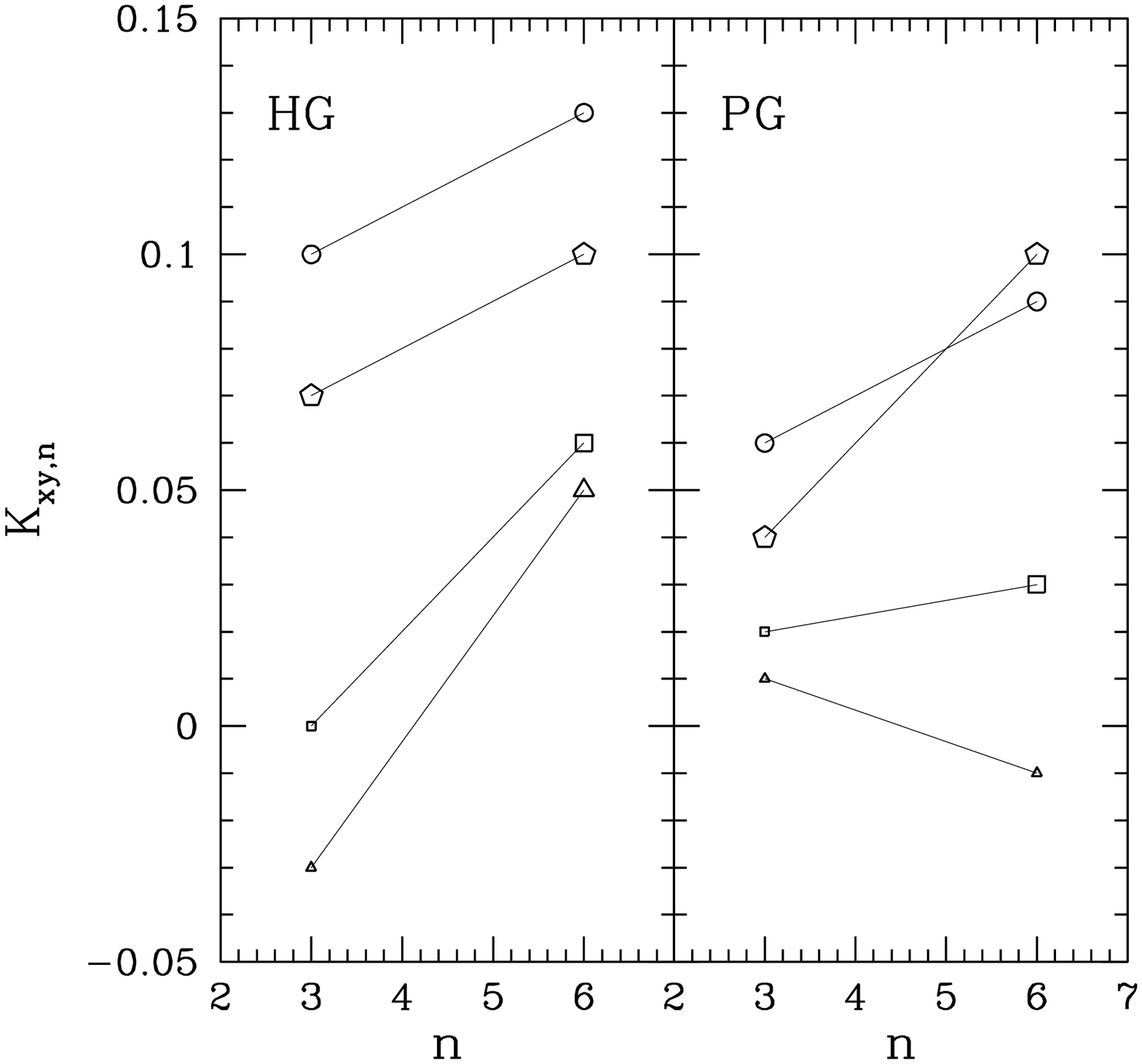}
\caption{Values of Kendall--correlation coefficients $K_{xy,n}$ for
poor $(n< 5)$ and rich $(n\ge 5)$ groups ($n=3$ and 6 are
the median values for poor and rich groups, respectively).  The
results for all four relations are shown: $R$--$M$ (circles), $R$--$T$
(pentagons), $V$--$M$ (squares), and $V$--$T$ (triangles).  Larger
symbols indicate a correlation with significance $\ge 90\%$.
}
\label{fig5}
\end{figure}

As for poor groups, they show fainter spatial segregation effects than
rich groups and no kinematical segregation effects at all (cf. also
Fig.~\ref{fig5}). This result could be due 1) to the dilution effect
of a significant number of spurious groups or 2) to some physical
difference between poor and rich groups. The most relevant difference
is probably the $\sigma_{\mathrm{v}}$, which is, on average, smaller
in poor than in rich groups (cf. Table~1).  However, segregation
effects are poorly or not dependent on $\sigma_{\mathrm{v}}$
(cf. Sect.~4.5); thus we are inclined to believe in the first
hypothesis.

Hereafter, we consider only more reliable, rich $n\ge5$ groups.

\subsection{Luminosity vs. morphology effects}

It is well known that galaxies of different morphological types have
different luminosity functions (e.g., Sandage et al.  \cite{san85};
Marzke et al. \cite{mar98}).  In particular, very late--type galaxies
strongly differ from other types having typically fainter magnitudes
(e.g., Sandage et al. \cite{san85}; Sandage \cite{san00}).  In our
case, morphological type $T$ and normalized magnitude $m-m_3$
correlate at the $98.8\%$ and $97.8\%$ c.l. for HG and PG,
respectively.

The problem of the independence of morphology and luminosity
segregations can be addressed by methods of partial correlation
(cf. Sect.~4.2). Thus, for the four relations considered in
Table~\ref{tab5} we estimate the Kendall partial rank correlation
coefficient considering the effect of luminosity (morphology) in the
relations involving morphology (luminosity).  Table~\ref{tab5} lists
the values of $K_{xy,n,m-m_3}$ and $K_{xy,n,T}$, and the respective
c.l.: they are similar to or just slightly smaller than the values of
corresponding $K_{xy,n}$. We conclude that morphology and
luminosity segregations are two independent effects.

Moreover, we consider early-- and late--type galaxies separately
(cf. Table~\ref{tab6}).  The $R$-$M$ correlation is significant for both
early-- and late--type samples. The same is true for the $V$-$M$
correlation in the HG case, but not in the PG case, where only the
early--type sample shows a significant $V$-$M$ correlation.

 \begin{table}
      \caption[]{Luminosity vs. morphology segregation}
         \label{tab6}
%
%
\begin{tabular}{crrrr}
\hline \hline
\multicolumn{1}{c}{$x$~vs.~$y$}
&\multicolumn{1}{c}{$N_{\rm{GALs}}$} 
&\multicolumn{1}{c}{$K_{xy,n}$, $P_{xy,n}$} 
&\multicolumn{1}{c}{$N_{\rm{GALs}}$}
&\multicolumn{1}{c}{$K_{xy,n}$, $P_{xy,n}$} 
\\
\hline
\multicolumn{1}{c}{}
&\multicolumn{2}{c}{HG Early--Types}
&\multicolumn{2}{c}{HG Late-Types}
\\
$R$--$M$&267&0.10, $\phantom{>} 99.7$&726&0.13, $>$99.9\\          
$V$--$M$&267&0.07, $\phantom{>} 96.9$   &726&0.05, $\phantom{>} 98.4$   \\      
\hline
\multicolumn{1}{c}{}
&\multicolumn{2}{c}{PG Early--Types}
&\multicolumn{2}{c}{PG Late-Types}
\\
\hline
$R$--$M$&344&0.08, $\phantom{>} 98.6$   &819&0.08, $>$99.9\\     
$V$--$M$&344&0.10, $\phantom{>} 99.9$&819&$-0.01$, $\phantom{>}  
63.5$\\      
\hline
\end{tabular}

   \end{table}

\subsection{Segregation and velocity dispersion}

The price to be paid for the large gain in statistics when using the
above ensemble groups is that we average away possible distinctive
behaviors. In particular, in the NOG sample, the value of velocity
dispersion ranges from very low--$\sigma_{\mathrm{v}}$ systems up to systems with
$\sigma_{\mathrm{v}} \sim 350$--$500$ \ks (cf. Fig.~\ref{fig1}).

\begin{figure}
   \centering
   \includegraphics[width=8cm]{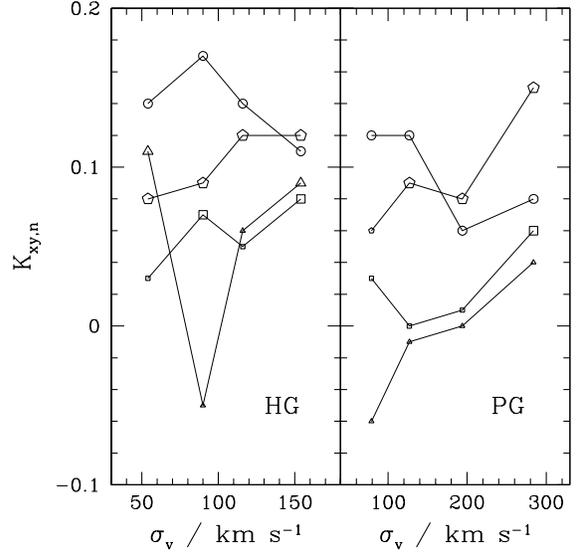}
\caption{Values of Kendall--correlation coefficients $K_{xy,n}$ as a
function of $\sigma_{\mathrm{v}}$ (here the median values in each quartile,
cf. Fig.~\ref{fig1}).  The results for all four relations are shown:
$R$--$M$ (circles), $R$--$T$ (pentagons), $V$--$M$ (squares), and
$V$--$T$ (triangles).  Larger symbols indicate a correlation with significance 
$\ge 90\%$.
\label{fig6}}
\end{figure}

We consider the four quartiles of the distribution of
$\sigma_{\mathrm{v}}$--values, i.e. each with $\sim 37$ and $\sim 42$
groups for HG5 and PG5, respectively.  The related ensemble systems
contain 184--341 galaxies for HG5 and 223--371 galaxies for PG5.
Fig.~\ref{fig6} shows the resulting Kendall coefficient for these
ensemble systems.  No systematic, general evidence of a connection
between segregation properties and velocity dispersion is shown.

\section{Summary \& Discussion}

We analyze $\sim$2x400 loose groups identified in the NOG catalog by
using hierarchical and ``friends--of--friends'' percolation catalogs
by G00.

Analyzing both catalogs we find similar results: earlier--type
(brighter) galaxies are more clustered and lie closer to the group
centers, both in position and in velocity, than later--type (fainter)
galaxies.  Spatial segregations are stronger than kinematical
segregations.  These effects are generally detected at the $\gtrsim$
3--sigma level, with the exception of morphological segregation in
velocity (cf. Sects.~3 and 4.2).  The significance of the last effect
is generally lower than $99\%$ in the comparison between weighted and
unweighted properties (cf. Sect.~3), and it is confirmed only for the
HG rich ensemble--group (cf. Sect.~4.3).

Our main results are confirmed by the analysis of statistically more
reliable $\sim 2\times 150$ rich groups (with at least five members,
cf. Sect.~4.3). Results for poorer groups are possibly diluted by
spurious systems.

The evidence of spatial morphology--segregation confirms previous
results recovered by using directly morphology data or morphological
indicators such as spectral types and colors (e.g., Postman \& Geller
\cite{pos84}; Mahdavi et al. \cite{mah99}; Tran et al. \cite{tra01};
Carlberg et al. \cite{car01b}; Dom\'{\i}nguez et al. \cite{dom02}).
The other kinds of segregation we detect 
are still poorly analyzed or debated questions for groups,
but are studied in the context of galaxy clusters.

Recent observational efforts to increase the number of members
of a few loose groups suggest that they are quasi--virialized system,
at least in their central regions, and show a continuum of properties
with respect to clusters, e.g., regarding the density profile of
galaxy distribution, the behavior of velocity dispersion profiles, and
the presence of substructures (Mahdavi et al. \cite{mah99}; Zabludoff
\& Mulchaey \cite{zab98a}, \cite{zab98b}).  In this context, it is
worth to attempt a more detailed comparison between our results for
groups and those obtained for clusters.

\begin{figure}
   \centering
   \includegraphics[width=8cm]{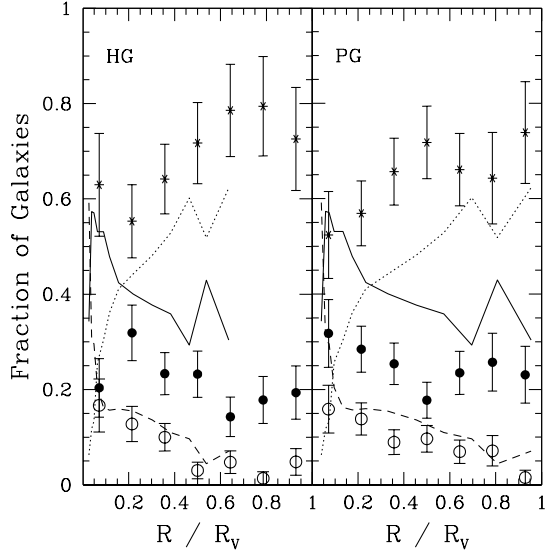}
\caption{Fraction of different morphological--types as a function of
distance from the group center: ellipticals (open circles),
lenticulars (filled circles), and spirals+irregulars (stars).  Points
are biweight mean values with 68\% Poissonian error--bars.
Results for rich $n\ge 5$ groups are shown. 
Whitmore et al. \cite{whi93} data for ellipticals (dashed line),
lenticulars (solid line), and spirals+irregulars (dotted line)
are also shown for a  qualitative comparison only
(see text).
\label{fig7}}
\end{figure}

In Fig.~\ref{fig7} we plot the relative fraction of ellipticals
($T<-3.5$), lenticulars ($-3.5\le T<0$), and spirals+irregulars ($T\ge
0$) for ensemble systems constructed from rich groups ($n\ge5$).  The
qualitative behavior resembles that already found for clusters
(e.g. Whitmore et al. \cite{whi93}): the fractions of ellipticals and
lenticulars decline with radius and the fraction of late--type
galaxies increases. For the sake of completeness we also plot cluster
data by Whitmore et al. which correspond to $\sim 2 R_{\mathrm{vir}}$
(i.e.  $\sim 1$ and $2/3$ $R_{\rm{V}}$ for PG and HG, respectively,
cf. Sect.~4.1).  Due to the different depth of the catalogs, the
possible inconsistencies in the morphological classification criteria,
and the degree of uncertainty in the normalization radius we do not
attempt a more quantitative comparison.

\begin{figure}
   \centering
   \includegraphics[width=8cm]{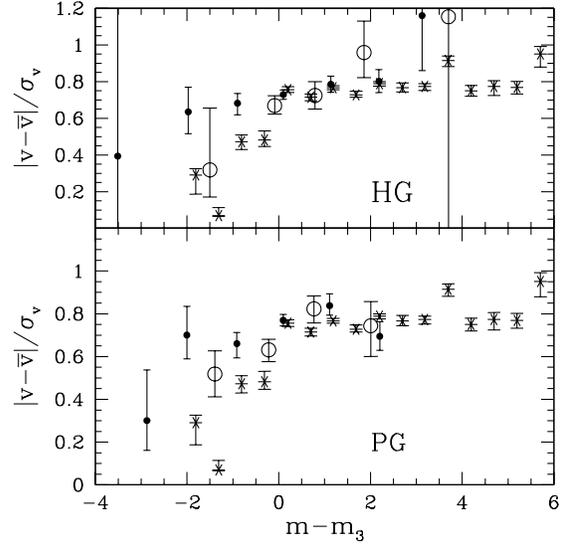}
\caption{$V$--$M$ relation of rich groups for all galaxies (filled
circles) and only for early--type galaxies (open circles).  It is
compared to the $V$--$M$ relation of clusters (stars, Biviano
et al. \cite{biv92}).
Points are biweight mean values with 68\% bootstrap error--bars.
Observational results for groups are normalized point by point with
results from simulated groups as in Fig.~\ref{fig3} and then
rescaled to the average value obtained from simulations.
\label{fig8}}
\end{figure}

In agreement with cluster studies (e.g., Biviano et al. \cite{biv02}),
we find significant evidence of both spatial and kinematical
segregation of galaxies of different luminosity, and luminosity
segregation seems to be related only to the most luminous galaxies.
Fig.~\ref{fig8} directly compares the $V$-$M$ relation for NOG groups
to that for clusters by Biviano et al. (\cite{biv92}) who found that
the $V$--$M$ relation first rapidly increases and then flattens out at
faint magnitudes.  According to Biviano et al., this trend can be
explained if galaxies brighter than $m_3$ have achieved the
energy--equipartition status, maybe due to dynamical friction or
galaxy merging, while fainter galaxies still lie in the
velocity--equipartition status generated by violent relaxation.  NOG
groups show a less sharp increase in the region of bright galaxies
with respect to clusters. A somewhat better agreement is obtained when
considering only early--type galaxies, which might better represent
the typical morphological content of those clusters. Another
explanation is that the removing of late--type galaxies reduces the
number of possible non--member galaxies highlighting better the true
physical relation.  Deeper samples for groups would be needed to
verify the flatness of the relation at faint magnitudes.

In NOG groups we find that luminosity segregation is found independent
of morphological segregation in agreement with the results from
clusters (e.g., Adami et al.  \cite{ada98}; Biviano et
al. \cite{biv02}).  In clusters luminosity segregation is evident only
for ellipticals and possibly lenticulars (e.g., Stein \cite{ste97};
Biviano et al. \cite{biv02}). In groups we do not reach a definitive
conclusion since we find that spatial luminosity segregation concerns
both early-- and late--type galaxies, while kinematical luminosity
segregation seems confined to early--type galaxies (in the case of PG
groups, cf.  Sect.~4.4).

Recent results suggest that morphological segregation in space
characterizes only massive groups, above some threshold value.  In
fact, Carlberg et al. (\cite{car01b}) found the presence of a color
gradient only in groups with $\sigma_{\mathrm{v}}>150$ \ks, and
Dom\'{\i}nguez et al. (\cite{dom02}) detected spectral--type
segregation only in groups with mass ${\cal M}\gtrsim 10^{13.5}$
\msun.  Additional support to this idea is given by our analysis in
Sect.~4.5, where the $R$-$T$ relation is strengthened with increasing
velocity dispersion (cf. pentagons in Fig.~\ref{fig6}).  However, no
systematic, general evidence of a connection between segregation
properties and velocity dispersion is shown and the question merits
further investigations.

To sum up, as regards coarse aspects of morphology and luminosity
segregation, our results are consistent with a continuum of properties
of galaxies in systems, from low--mass groups to massive clusters.
This result is in agreement with the early study by Postman \& Geller
(\cite{pos84}) on morphological segregation and, e.g., more recent
results by Lewis et al. (\cite{lew02}) who found that environmental
influences on galaxy star formation are not restricted to cluster
cores, but are effective in groups, too.

Our results suggest that the segregation effects we analyze are mainly
connected with the initial conditions at the time of galaxy formation.
Alternatively, the mechanisms which influence galaxy luminosity and
morphology should act in a similar way in groups and in clusters, or,
at least, in the subunits forming clusters in the context of the
hierarchical scenario.

In this framework, it is also worth discussing our results in
connection to those coming from large--scale analyses, since a large
fraction of galaxies is located in groups ($\sim 40\%$, cf. Ramella et
al. \cite{ram02}).  Spatial segregation phenomena are generally
studied through the galaxy--galaxy correlation function analysis.
This kind of analysis is fully independent and complementary to our
work, which concerns very small scales. As for NOG catalog, Giuricin
et al. (\cite{giu01}) analyzed the redshift--space two--point
correlation function of galaxies, finding that -- on scales between
$\sim 3$ and 10--20 \h -- the strength of clustering is stronger for
earlier and, independently, for more luminous galaxies, in agreement with
most recent literature (cf. Giuricin et al.  and refs. therein).  The
projected correlation function overcomes the problem of distortions of
the clustering pattern induced by peculiar motions and allows to
perform analysis down to smaller scales ($\lesssim 0.5$ \h). In
particular, very recent studies of the 2dF Galaxy Redshift Survey show
that the strength of the clustering of luminous galaxies increases
with the galaxy luminosity and that luminosity and spectral--type
segregations are independent effects (Norberg et al. \cite{nor01};
\cite{nor02}).  Thus, segregation phenomena, qualitatively in
agreement with those detected in our study, are present out to very
large scales. This could be another piece of evidence in favor of a
mostly primordial origin for spatial segregation effects.

\begin{acknowledgements}
We thank Andrea Biviano, Manuela Magliocchetti, Christian Marinoni,
and Massimo Ramella for useful suggestions and discussions. We
also thank an anonymous referee for a careful reading of the
manuscript and suggestions which improved the paper.  Work partially
supported by the Italian Ministry of Education, University, and
Research (MIUR, grant COFIN2001028932 "Clusters and groups of
galaxies, the interplay of dark and baryonic matter"), and by the
Italian Space Agency (ASI).
\end{acknowledgements}

\end{document}